\documentclass[lettersize,journal]{IEEEtran}
\usepackage{amsmath,amsfonts, bm}
\usepackage{algorithmic}
\usepackage{algorithm}
\usepackage{array}
\usepackage[caption=false,font=normalsize,labelfont=sf,textfont=sf]{subfig}
\usepackage{textcomp}
\usepackage{stfloats}
\usepackage{url}
\usepackage{verbatim}
\usepackage{graphicx}
\usepackage{cite}
\usepackage{physics}
\usepackage {tabularx} 
\usepackage{amssymb}
\usepackage{amsmath}
\usepackage[]{units}
\usepackage[]{upgreek}
\begin{document}
\title{ Amplification and frequency conversion of spin waves using acoustic waves}

\author{Morteza Mohseni, Abbass Hamadeh, Moritz Geilen 
and  Philipp Pirro
\thanks{M. Mohseni, A. Hamadeh (corresponding author), M. Geilen and  P. Pirro (corresponding author) are with Fachbereich Physik and Landesforschungszentrum OPTIMAS, Rheinland-Pf\"alzische Technische Universit\"at Kaiserslautern-Landau, 67663 Kaiserslautern, Germany (e-mail: hamadeh@rhrk.uni-kl.de)}
}



\maketitle

\begin{abstract}
We numerically study the acoustic parametric amplification of spin waves using surface acoustic waves (SAW) in a magnetic thin film. First, we illustrate how the process of parametric spin-wave generation using short-waved SAWs with a fixed frequency allows to tune frequencies of the generated spin waves by the applied magnetic field. We further present the amplification of microwave driven spin waves using this method. The decay length and the amplitude of the driven spin waves can be amplified up to approximately 2.5 and 10 times compared to the reference signal, respectively. More importantly, the proposed design can be used as a frequency converter, in which a low (high) frequency spin-wave mode stimulates the excitation of a high (low) frequency mode. Our results pave the way in designing highly flexible and efficient hybrid magnonic device architectures for microwave data transport and processing.  
\end{abstract}


\section{Introduction}
\IEEEPARstart{W}{ave-based }computing technologies aim to propose alternative solutions for data transport and processing in device architectures, as referenced in various studies (e.g. \cite{barman20212021,mahmoud20204,mahmoud2021spin,chumak2022advances,chumak2019fundamentals}). These data processing units have the potential to complement current CMOS technologies as they can use the frequency and phase of waves as new degrees of freedom for data manipulation \cite{pirro2021advances}.

Among classical waves, spin waves (SWs) are very attractive candidates for various applications due to several reasons. For instance, they lack heat dissipation due to the absence of Joule heating losses, particularly in insulators \cite{serga2010yig,khitun2010magnonic,chumak2015magnon}. They also possess strong inherent nonlinearity \cite{gurevich2020magnetization} , small footprints down to nanometers \cite{talmelli2021electrical,zografos2015design}, and can potentially be integrated with current microwave interface technologies such as surface acoustic waves (SAWs) and bulk acoustic waves (BAWs) \cite{kittel1958interaction}.

Several data processing devices using spin waves have already been proposed or realized, including transistors \cite{chumak2014magnon,chumak2017magnonic}, directional couplers \cite{wang2020magnonic}, amplifiers \cite{divinskiy2018excitation,bracher2017parallel}, majority gates \cite{sadovnikov2016frequency,fischer2017experimental,talmelli2020reconfigurable} , and many more \cite{gutierrez2017magnonic,rahman2015wave,hamadeh2022hybrid,baumgaertl2022reversal}. However, one  challenging issue facing these technologies is the inherent magnetic losses which prevent long-distance propagation and, consequently, result in a short decay length \cite{eshbach1960surface}. This hinders the design of universal magnonic elements with all the necessary functionalities, in which the information can be processed completely in the magnonic domain \cite{wang2018reconfigurable}. In addition, the design of coherent wave-based logic often relies on the use of a specific frequency range for a given device to provide the desired functionality. For example, the gate magnons in the magnon-scattering based magnonic transistor \cite{chumak2014magnon} need to have a different frequency than the signal magnons. The use of these two different frequencies significantly is a challenge for the cascading of magnon transistors. Thus, to connect different magnonic devices, a way to transform the signal-carrying spin-wave frequency would open new opportunities for more efficient and flexible designs of magnonic circuits.

Both issues, the compensation of damping and the frequency transformation, can be overcome by the use of parametric processes. The most established method is the parallel parametric amplification of spin waves\cite{mohseni2020parametric, heinz2022parametric,bracher2017parallel,Melkov2001,Melkov1999,Chowdhury2017}. Parametric amplification of spin waves is based on the splitting of a pumping wave (e.g. microwave photon or phonon) into two counter-propagating magnons with half of the pumping frequency. It has several advantages compared to other methods, e.g.the amplification via spin transfer torque\cite{bracher2017parallel,Gladii2016,Evelt2016}. For example, it is phase-sensitive \cite{breitkreutz2016design,Bracher2016}, it covers a large range of wave vectors \cite{verba2018amplification,Sandweg2011}. When using magneto-elastic fields \cite{thevenard2014surface,weiler2011elastically,verba2018nonreciprocal} or voltage-controlled magnetic anisotropy \cite{Verba2014} it can be highly energy-efficient \cite{Chowdhury2017,Lisenkov2019}. In this case, the magneto-elastic coupling between phonons and magnons is used to excite magnons parametrically. This effect, known as acoustic pumping of SWs, can use SAW elements, a well-established device in many modern microwave technologies, to drive SWs to very high amplitudes where nonlinear effects such as parametric instabilities play a major and interesting role \cite{keshtgar2014acoustic}.

Recently, we experimentally demonstrated the acoustic pumping of SWs from the thermal bath using SAWs in a hybrid magneto-acoustic element \cite{geilen2022parametric}. Here, we extend our studies and show how an optimized geometry leads to a parametric generation of magnons which is not disturbed by off-resonant excitations. This allows to easily tune the frequency and wave vector of the parametrically pumped magnons by changing the applied magnetic bias field while keeping the SAW frequency fixed.
 Moreover, we demonstrate that a coherently excited SW packet can be amplified using SAWs. By driving the system with a SAW amplitude below the instability threshold, an amplification of the local SW amplitude up to 2.5 times has be achieved. Additionally, we show that the proposed device can be used as a frequency converter, in which the input of a low (high) frequency SW mode leads to the generation of a high (low) frequency SW mode of well defined frequency.

\section{Device Design}

\begin{figure}[!t]
\centering
\includegraphics[width=3.2in]{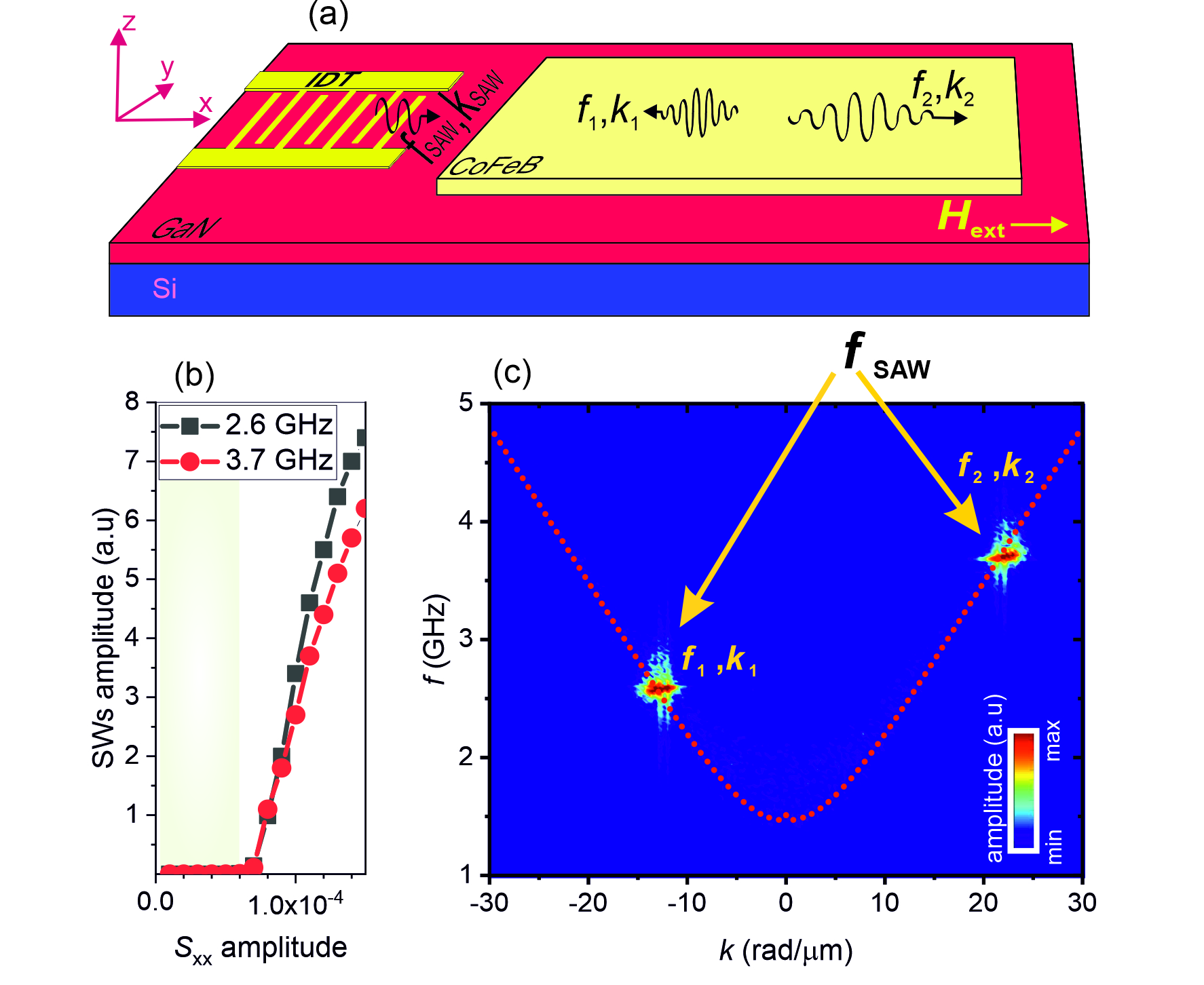}
\caption{(a) Schematic representation of the proposed device; (b) threshold curve of the parametric generation of spin waves using surface acoustic waves; and (c) analytically calculateed spin-wave dispersion relation(dashed red line) for the investigated  magnetic film in which the spin waves are parametrically generated using surface acoustic waves. The color plot shows the simulation results demonstrating the generation of two magnon modes fulfilling the energy can momentum conservation laws for a splitting of a SAW phonon.}
\label{fig1}
\end{figure}

The proposed device is schematically presented in Fig.~\ref{fig1}a. It consists of an interdigital transducer (IDT) on top of a GaN/Si stack. The IDT has a finger-to-finger distance of \unit[170]{nm}, which leads to the excitation of SAWs with a wavelength of $\lambda_\mathrm{SAW}$ = \unit[680]{nm} and a wave vector of $k_\mathrm{SAW}$ = \unit[9.2]{rad/$\mu m$} at the first harmonic resonance frequency, equal to $f_\mathrm{SAW}$ = \unit[6.3]{GHz}. This model is based on previously realized structures. Further details about the experimental investigation of SAWs in this system can be found in Ref. \cite{geilen2020interference}. The magnetic part of the device is a \unit[20]{nm} thick CoFeB thin film which is placed on top of the substrate stack adjacent to the IDT. In contrast to previous studies \cite{geilen2022fully}, the external field is applied along the x-direction of the system, so the dominant strain component $S_{xx}$ of a Rayleigh-like SAW travelling in x direction is not exerting any torque on the static magnetisation. 

 The micromagnetic simulations have been carried out using the open source MuMax 3.0 package \cite{mumax3} along with the Aithericon software platform \cite{aithericon}. The following parameters which resembles values from the experimental evaluations are used in the simulations \cite{geilen2022fully}: $M_\mathrm{S}=\unit[950]{kA/m}$, $A_\mathrm{ex}=\unit[15]{pJ/m}$, $\alpha=\unit[0.004]{}$, $B_1=\unit[9.38]{MJ/m^3}$ and $B_2=\unit[9.38]{MJ/m^3}$. The simulated system is a pad with the dimensions of \unit[20]{$\mu \mathrm{m}$} $\times$ \unit[2]{$\mu \mathrm{m}$} $\times$ \unit[20]{nm} which is divided into 2048 $\times$ 128 $\times$ 1 cells. We have used periodic boundary conditions in both x and y directions to mimic a plane film. The SAW with wave vector $k_\mathrm{SAW}$= \unit[9.2]{rad/$\mu m$} and frequency $f_\mathrm{SAW}=\unit[6.3]{GHz}$ is represented by a plane wave with a certain amplitude for the strain component $S_{xx}$. As we will show below, parametric generation of spin waves using SAW is possible in such a condition. The simulation are carried out assuming room temperature T = \unit[300]{K}. The dynamic components of the magnetization are collected and analyzed using a fast Fourier transformation in space and time. Further details about data analysis can be found in Ref.\cite{geilen2022parametric}.

The parametric generation of SWs is a nonlinear effect that requires a certain level of input energy, known as the parametric threshold. The threshold is affected by factors such as magnetic losses and the interaction between magnons and the pumping field, which in this study is generated by the surface acoustic wave. The process of generating SWs through SAW-based pumping can be described using a rate equation for the SW amplitudes, $c_k$ \cite{Lisenkov2019}:

\begin{equation}
\label{deqn_eq1}
\frac{d\mathrm{c}_1}{dt} + i\omega_1\mathrm{c}_1 + \Gamma_1\mathrm{c}_1 = iV_{\mathrm{SAW},12}q_{\mathrm{SAW}}\mathrm{c}_2^* \quad
\end{equation}

\begin{equation}
\label{deqn_eq2}
V_{\mathrm{SAW},12} = \frac{\gamma B_1}{M_s}\overline{S}\left[m_{1,x}^*m_{2,x}^* - \bm{m}_1^*\cdot \bm{m}_2^*\cos^2\left(\varphi\right)\right] \quad,
\end{equation}

where $q_{\mathrm{SAW}}$ is the SAW amplitude, $V_{(\mathrm{SAW},12)}$ is the parametric coupling efficiency, $\cos\left(\varphi\right)$ is the angle between the static magnetization and the propagation direction of the SAW, $\overline{S}_{0,xx}$ measures the thickness-averaged SAW strain, and $\bm{m}_k$ is the SW mode structure. Since the translation symmetry is not broken in the investigated system it is important to note that this process is also wave-vector conserving, which is fulfilled when $k_{\mathrm{SAW}} = k_1 + k_2$. Energy conservation implies $f_{\mathrm{SAW}} = f_1 + f_2$.

To simulate the process, we fix the magnetic field to $\mu_0\mathrm{Hext\ }= 2,\mathrm{mT}$ pointing along the wave vector of the SAW ($\phi=0$) as shown in Fig. \ref{fig1}b. The threshold for parametric magnon generation is observed above a strain amplitude $S_{xx}$ of $6\times10^{-5}$. The pumped magnons populate two spectral positions of the fundamental SW mode while addressing the energy and momentum conservation laws $(f_1 = 2.6,\mathrm{GHz}, k_2 = -12.5,\mathrm{rad/}\mu\mathrm{m}\mathrm{\ and\ }f_2 = 3.7,\mathrm{GHz}, k_2 = 21.7,\mathrm{rad/}\mu\mathrm{m})$, as shown in Fig. \ref{fig1}c.

\section{Results and Discussion}

\begin{figure}[!t]
\centering
\includegraphics[width=3.2in]{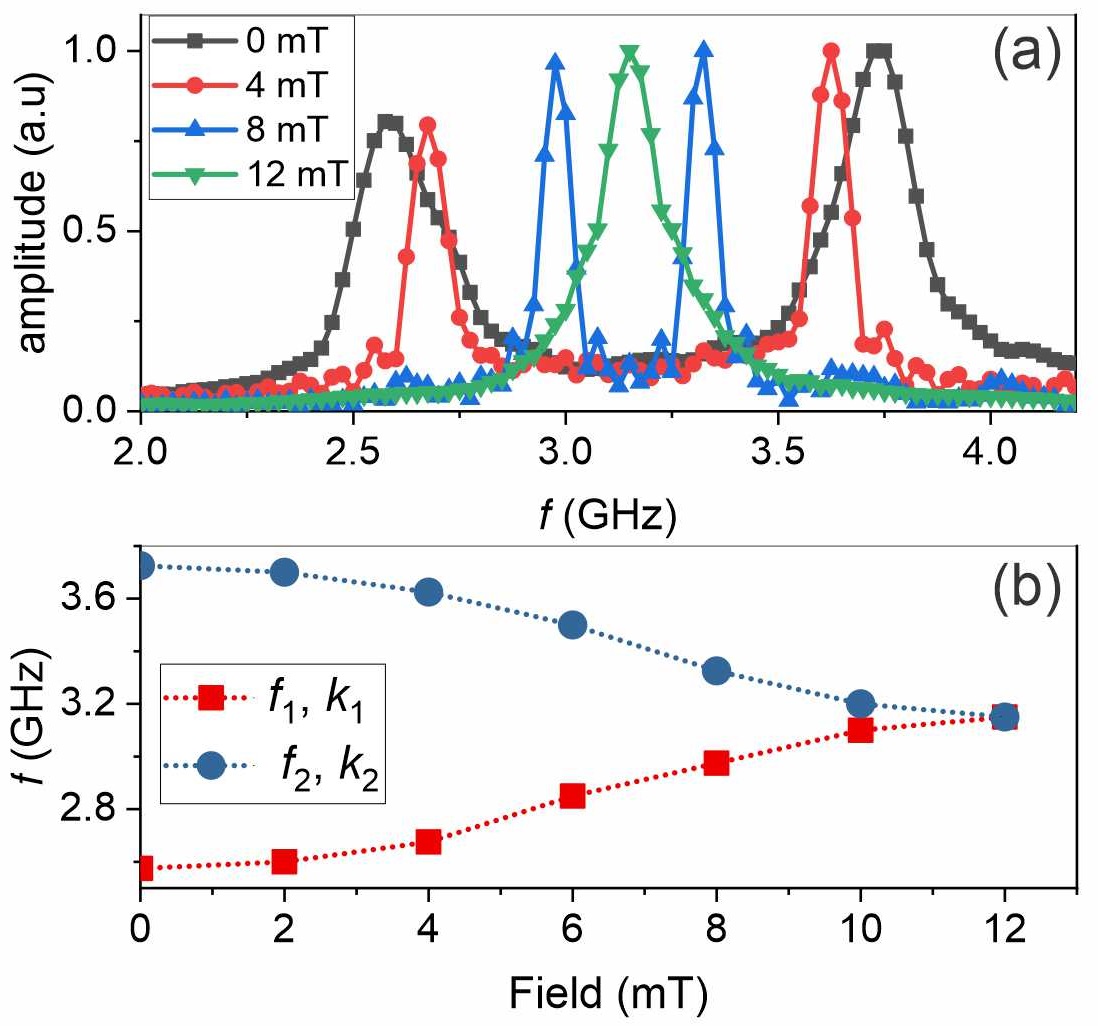}
\caption{(a-b) The frequency spectra of the parametrically generated spin waves in the presence of various magnetic fields; b) parametrically generated spin waves amplitude as a function of external magnetic field taken from (a).}
\label{fig2}
\end{figure}

The investigation of the frequency tunability of the SWs generated by acoustic pumping using an external magnetic field is presented in Fig. \ref{fig2}a. It shows the frequency spectra of the pumped magnons for different magnetic fields. As a result of conservation of energy and momentum and the field-dependent spin-wave dispersion relation,  the frequency of the pumped pairs of magnons $(f_1, k_1)$ and $(f_2, k_2)$ is changing and approaches $f_{\mathrm{SAW}}/2$ as the magnetic field increases. This behavior is further highlighted in Fig.~\ref{fig2}b which shows the frequency of the pumped magnons as a function of the applied field. At and above a field of $\mu_0\mathrm{Hext\ }$= \unit[12]{$\mathrm{mT}$}, magnons are pumped below the SW spectrum due to non-resonant pumping process \cite{mohseni2020parametric}.

In the following, and in order to study the amplification and frequency conversion of SWs using SAW, we will keep the amplitude of the SAW in the sub-threshold range as shown by the shaded area in Fig. \ref{fig1}b. This prevents the generation of magnons starting from the thermal level.

\begin{figure}[!t]
\centering
\includegraphics[width=3.2in]{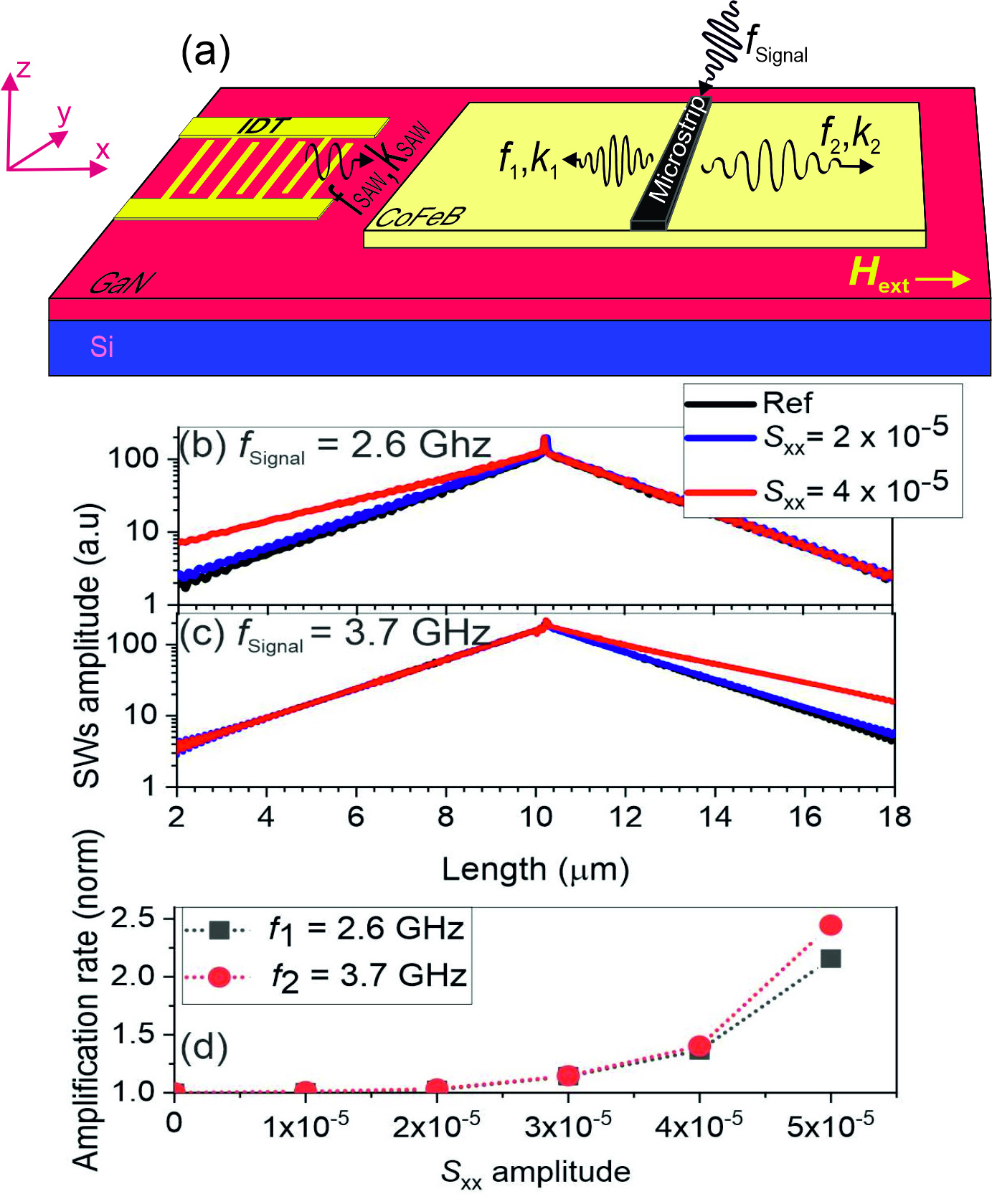}
\caption{a) Schematic representation of the proposed device including a microstrip antenna, b-c) Spin-wave amplitude as a function of propagation distance in the presence surface acoustic waves with different amplitudes while the driven mode has a fixed frequency; c) the decay length amplification rate as a function of surface acoustic wave amplitude.}
\label{fig3}
\end{figure}

We now study the amplification of the externally excited SWs using the SAW. To this end, we inject a SW packet (with a duration of \unit[5]{ns}) having a frequency of  $f_{\mathrm{Signal}}$ 
which is either equal to $f_1$ or $f_2$. For this purpose as shown in Fig.\ref{fig3}a, a microstrip antenna is simulated on top of the CoFeB in order to excite SW packets resonantly.

 Fig.~\ref{fig3}b show the decay characteristics of the driven SWs with the frequency of $f_{\mathrm{Signal}}$ = $f_1 = \unit[2.6]{GHz}$ while the amplitude of the $S_{xx}$ varies from 0 (reference signal) to $2\times10^{-5}$ (blue curve) and  $4\times10^{-5}$ (red curve). Obviously, the decay rate of the SWs has increased from $d_{\mathrm{signal}} = 1.95\,\mu\mathrm{m} \mathrm{\ to\ } d_{\mathrm{amplified}} = 2.69\,\mu\mathrm{m}$ which corresponds to an amplification factor of \unit[1.37]{}. Please note that such an amplification is nonreciprocal ($f_1$ has a negative k-vector) due to the wave vector conservation. Changing the frequency to $f_{\mathrm{Signal}}$ = $f_2$ = \unit[3.7]{GHz}, leads to a similar amplification as shown in Fig.~\ref{fig3}c. However, due to the non-reciprocity of the pumping effect using SAW, at this frequency only $+k$ magnons are amplified. Fig.~\ref{fig3}d summarizes the normalized (to the reference signal) amplification factor of the decay length as a function of the SAW amplitude. Evidently, an amplification of up to \unit[2.5]{} times can be achieved. At higher strain amplitudes the parametric generation of SWs from the thermal level occurs which hinders an efficient amplification of the signal wave \cite{bracher2017parallel}.

\begin{figure}[!t]
\centering
\includegraphics[width=3.2in]{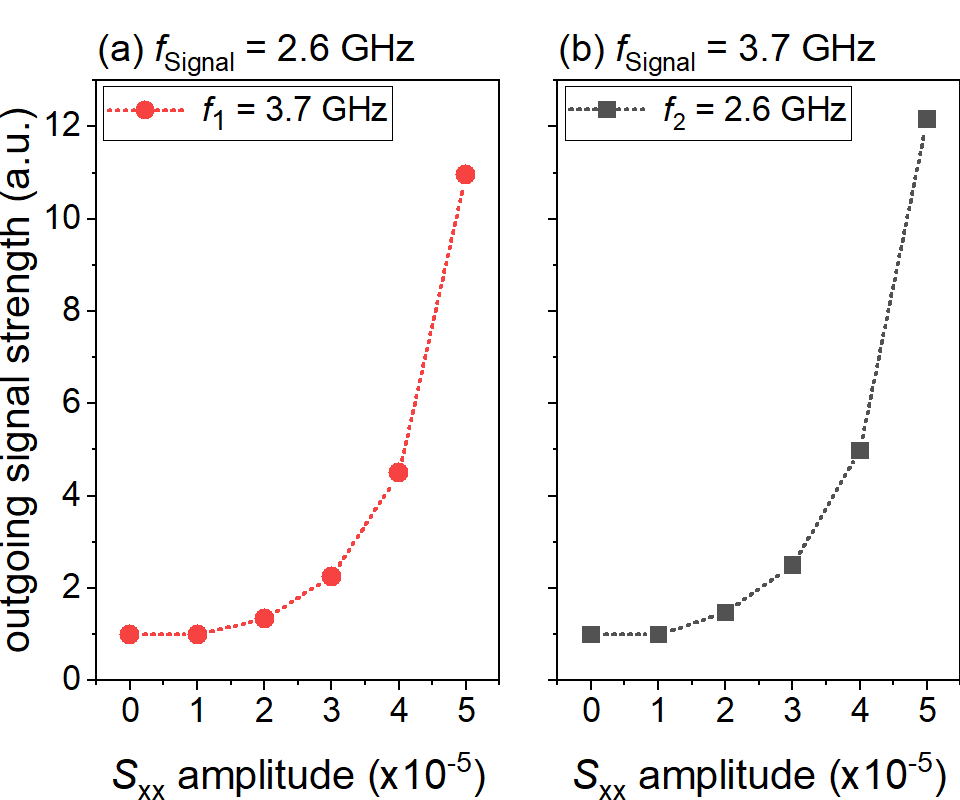}
\caption{Outgoing signal strength of spin waves with respect to the reference signal (thermal level) as a function of surface acoustic wave amplitude for two frequencies  of $f_{\mathrm{Signal}}$ : (a) \unit[2.6]{GHz} and (b) \unit[3.7]{GHz}.}
\label{fig4}
\end{figure}

Due to the conservation of the energy and momenta during the parametric amplification, the pumped magnons always are generated in pairs. Due to the finite momentum of the pumping wave, the usual degeneracy of the parametrically amplified waves is canceled. When the SAW amplifies SWs e.g. at $-|k_1|$, the paired magnons at $+|k_2|$ must be amplified from the thermal bath as well.  Such an effect allows us to use the SAW-based amplifier as a frequency converter since driving the system with e.g. $f_{\mathrm{Signal}}$ = $f_1$ stimulates the magnon generation at $f_2$. We now analyze this effect in more details. Figure \ref{fig4}a illustrates the outgoing signal strength of the paired spin waves (SW) at \unit[3.6]{GHz}  as a function of surface acoustic wave (SAW) amplitude (which is proportional to the strain $S_{xx}$) when a SW packet with frequency  of $f_{\mathrm{Signal}}$ = \unit[2.6]{GHz} is injected. Evidently, 
we can see that the paired magnons at $f_2$ = \unit[3.6]{GHz} exhibit a steep increase with increasing strain amplitude. Figure \ref{fig3}b presents the same analysis when $f_{\mathrm{Signal}}$ = \unit[3.7]{GHz}, where the amplitude of the paired magnons at $f_2$ = \unit[2.6]{GHz} can be as high as \unit[12]{times} relative to the thermal reference level. By chaining the magnetic field, the distance between the signal and the stimulated spin-wave can be adjusted, makes this process a valuable tool for frequency conversion and the inversion of the propagation direction of the frequency-converted signal.

\section{Conclusion}

In summary, our numerical study has revealed the potential for efficient generation and amplification of spin waves (SWs) in a magnetic film through the use of surface acoustic waves. We have demonstrated that this method can achieve a significant increase of the spin-wave decay length. In addition, the proposed mechanism offers frequency tunability via the magnetic field . The experimental realization of this structure layout has the potential to lead to the design of highly efficient magnonic elements, as the excitation of SAWs requires only electrical fields which significantly reduces Joule losses compared to the traditional driving of the magnonic system using the Oersted fields of RF currents. Last but not least, the demonstrated ability to convert the spin-wave frequency and the spin-wave wavelength opens a completely new design space for hybrid magnonic devices.

\section*{Acknowledgments}

This research was funded by the European Research Council within the Starting Grant No. 101042439 "CoSpiN" and by the Deutsche Forschungsgemeinschaft (DFG, German Research Foundation) within the Transregional Collaborative Research Center—TRR 173–268565370 “Spin + X” (project B01). We would like to extend our gratitude to Roman Verba from the Institute of Magnetism, Kyiv, Ukraine for his insightful discussions. 

%

\bibliographystyle{IEEEtran}
\bibliography{references}

\newpage

 




\vfill

\end{document}